\documentclass[aps,pra,superscriptaddress,showpacs,twocolumn]{revtex4}

\usepackage{amsmath}
\usepackage{amssymb}
\usepackage{graphicx}

\def\prl{Phys. Rev. Lett. }
\def\pra{Phys. Rev. A }
\def\<{\langle}
\def\>{\rangle}
\def\dket#1{| #1\>\!\>}

\def\dbra#1{\<\!\< #1|}

\def\dinner#1#2{\<\!\< #1| #2\>\!\>}

\def\douter#1#2{| #1\>\!\>\<\!\< #2|}

\def\norm#1{|\!|#1|\!|}

\begin{document}

\title{Quantum State Tomography with Joint SIC~POMs and Product SIC~POMs}
\author{Huangjun Zhu }
\affiliation{Centre for Quantum Technologies, %
National University of Singapore, Singapore 117543, Singapore}
\affiliation{NUS Graduate School for Integrative Sciences and
Engineering, Singapore 117597, Singapore}

\author{Berthold-Georg Englert}
\affiliation{Centre for Quantum Technologies, %
National University of Singapore, Singapore 117543, Singapore}
\affiliation{Department of Physics, %
National University of Singapore, Singapore 117542, Singapore}



\begin{abstract}

We introduce random matrix theory to study the tomographic
efficiency of  a wide class of measurements constructed out of
weighted 2-designs, including symmetric informationally complete
(SIC) probability operator measurements (POMs). In particular, we
derive analytic formulae for the mean Hilbert-Schmidt distance and
the mean trace distance between the estimator and the true state,
which clearly show the difference between the  scaling behaviors of
the two error measures with the dimension of the Hilbert space. We
then prove that the product SIC~POMs---the multipartite analogue of
the SIC~POMs---are optimal among all product measurements in the
same sense as the SIC~POMs are optimal among all joint measurements.
We further  show that, for bipartite systems, there is only a
marginal efficiency advantage of the joint SIC~POMs over the product
SIC~POMs. In marked contrast, for multipartite systems, the
efficiency advantage of the joint SIC~POMs increases exponentially
with the number of parties.
\end{abstract}

\date{\today}

\maketitle

\section{Introduction}
Quantum state tomography is a procedure for inferring the state of a
quantum system from generalized measurements. It is a primitive of
quantum computation, quantum communication, and quantum
cryptography, because all these tasks rely heavily on our ability to
determine the state of a quantum system at various stages. One of
the main challenges in quantum state tomography is to reconstruct a
generic unknown quantum state as efficiently as possible and to
determine the resources necessary to achieve a given accuracy, which
can be quantified by various figures of merit, such as the trace
distance, the Hilbert-Schmidt (HS) distance, or the fidelity
\cite{ParR04,LvoR09}.

A generalized measurement in quantum mechanics is known as a
probability operator measurement (POM). 
A measurement is \emph{informationally complete} (IC) if any state
is determined completely by the measurement statistics \cite{Pru77,
Bus91, DarPS04}. In a $d$-dimensional Hilbert space, an IC
measurement consists of at least $d^2$ outcomes, whereas  a
\emph{minimal} IC measurement consists  of no more than $d^2$
outcomes. A particularly appealing choice of IC measurements are
those constructed out of \emph{weighted 2-designs}, called
\emph{tight} IC measurements according to Scott \cite{Sco06,RoyS07}.
Under linear quantum state tomography, they not only feature  a
simple state reconstruction formula but also minimize the mean
squared error (MSE)--- the mean square HS distance between the
estimator and the true state \cite{Sco06}. The construction of tight
IC measurements has also been discussed in detail in
Ref.~\cite{RoyS07}.

A \emph{symmetric informationally complete} (SIC) POM  \cite{Zau11,
RenBSC04, App05, ScoG10} is a very special tight IC measurement
composed of $d^2$ subnormalized projectors onto pure states with
equal pairwise fidelity of $1/(d+1)$.  They may be considered as
fiducial measurements for state tomography for reasons of  their
high symmetry and high tomographic efficiency \cite{RenBSC04, App05,
ScoG10, RehEK04, Sco06}. It is widely believed that SIC~POMs exist
in any Hilbert space of finite dimension since Zauner posed the
conjecture \cite{Zau11}, although no rigorous proof is known.
Analytical solutions of SIC~POMs are known for $d\leq16$ and $d=19,
24, 28, 35, 48$, see Refs. \cite{Zau11, RenBSC04, App05, ScoG10,
Gra11} and the references therein; numerical solutions with high
precision have also been found up to $d=67$ \cite{RenBSC04,
RenBSC04b, ScoG10}. In addition to their significance in quantum
state tomography, SIC~POMs have attracted much attention due to
their connections with mutually unbiased bases (MUB)
\cite{Iva81,WooF89, Woo06, App09, DurEBZ10}, equiangular lines
\cite{LemS73}, Lie algebras \cite{AppFF11}, and foundational studies
\cite{Fuc10}.

The trace distance is one of the most important distance measures
and distinguishability measures in quantum mechanics, and is widely
used in quantum state tomography, quantum cryptography, and
entanglement theory \cite{ParR04, NieC00, BenZ06, FucG99, HorHHH09},
as well as other contexts. In addition, it is closely related to
other important figures of merit, such as the fidelity and the
Shannon distinguishability \cite{NieC00,FucG99}. However, little is
known about the tomographic resources required to achieve a given
accuracy as measured by the trace distance, since its definition
involves taking the square root of a positive operator. Even for the
qubit SIC~POM \cite{RehEK04,LinLK08, BurLDG08}, no analytic formula
is known concerning the mean trace distance between the estimator
and the true state. One motivation of the present study is to solve
this long-standing open problem.

In the case of a bipartite or multipartite system, it is
technologically much more challenging to perform full joint
measurements such as  SIC~POMs on the whole system. Moreover, in
some important realistic scenarios, such as  tomographic quantum key
distribution \cite{LiaKEK03, EngKRN04, DurKLL08}, all parties are
space separated from each other, and it is impractical to perform
full joint measurements. Nevertheless, each party can perform a
local SIC~POM and reconstruct the  global state after gathering all
the data obtained. Such a POM will henceforth be referred to as a
product SIC~POM; by contrast, the SIC~POM of the
whole system will be referred to as a joint SIC~POM. 
The product SIC~POM is particularly appealing in tomographic quantum
key distribution since it minimizes the redundant information and
classical communication required to exchange measurement data among
different parties \cite{DurKLL08}. However, even less is known
concerning the tomographic efficiency of the product SIC~POM except
for  numerical studies in the two-qubit  setting
\cite{BurLDG08,TeoZE10}.

In this article, we aim at characterizing the tomographic efficiency
of tight IC measurements in terms of the mean trace distance and the
mean HS distance, with special emphasis on the minimal tight IC
measurements---SIC~POMs. We also determine the efficiency gap
between the product measurements and the joint measurements  in the
bipartite and multipartite settings.

First, we introduce random matrix theory \cite{Meh04} to study the
tomographic efficiency of tight IC measurements and derive
analytical formulae for the mean trace distance and the mean HS
distance. We illustrate the general result with  SIC~POMs, and show
the different scaling behaviors of the two error measures with the
dimension of the Hilbert space. In the special case of the qubit
SIC~POM, we discuss in detail the dependence of the reconstruction
error on the Bloch vector of the unknown true state and make contact
with the experimental data given by Ling {\it et al.}
\cite{LinLK08}. As a by product, we also discovered  a special class
of tight IC measurements that feature exceptionally symmetric
outcome statistics and low fluctuation over repeated experiments.

Second, in the bipartite and multipartite settings, we show that the
product SIC~POMs are optimal among all product measurements in the
same sense as the joint SIC~POMs are optimal among all joint
measurements. We further show that for bipartite systems, there is
only a marginal efficiency advantage of the joint SIC~POMs over the
product SIC~POMs. Hence, it is not worth the trouble to perform the
joint measurements. However, for multipartite systems, the
efficiency advantage of the joint SIC~POMs increases exponentially
with the number of parties.

To provide a simple picture of the tomographic efficiency of
SIC~POMs and product SIC~POMs, we restrict our attention to the
scenario where the number of copies of the true states available is
large enough to yield a reasonably good estimator,  and we focus on
the standard state reconstruction scheme, also known as linear state
tomography \cite{ParR04,Sco06}. The analysis of the efficiencies of
other reconstruction schemes, such as the maximum likelihood method
\cite{ParR04, Hra97,TeoZE10}, is much more involved. Hopefully, our
analysis can  serve as a starting point and, in principle, it can be
generalized to deal with those more complicated situations.
Moreover, for minimal tomography on a large sample, the estimator
given by the standard reconstruction scheme is almost identical to
that given by the maximum-likelihood method, except when the true
state is very close to the boundary of the state space. Hence, the
efficiencies of the two alternative schemes are quite close to each
other in this scenario.

The rest of this article is organized as follows. In
Sec.~\ref{sec:general}, we review  the  basic framework for linear
state tomography and recall the concepts of weighted $t$-designs,
SIC~POMs and tight IC measurements. In
Sec.~\ref{sec:RandomMatrixST}, we introduce random matrix theory to
study the tomographic efficiency of tight IC measurements, in
particular the SIC~POMs. We derive analytical formulae for the mean
trace distance and the mean HS distance and illustrate the
difference in their scaling behaviors with the dimension of the
Hilbert space. In Sec.~\ref{sec:sicprod}, we first prove the
optimality of the product SIC~POMs among product measurements and
then compare the tomographic efficiencies of the product SIC~POMs
and the joint SIC~POMs. We conclude with a summary.

\section{\label{sec:general}Setting the stage}

\subsection{\label{sec:LinearST}Linear state tomography}
 A  generalized measurement is composed of a set of outcomes
represented mathematically by positive operators $\Pi_j$ that sum up
to the identity operator 1. Given an unknown true state $\rho$, the
probability of obtaining the outcome $\Pi_j$ is given by the Born
rule: $p_j=\mathrm{tr}(\Pi_j\rho)$. A measurement is IC if we can
reconstruct any state according to the statistics of measurement
results, that is the set of probabilities $p_j$.  If we  take both
the state $\rho$ and the outcomes $\Pi_j$ as vectors in the space of
Hermitian operators, then the probability can be expressed as an
inner product $\dinner{\Pi_j}{\rho}\equiv\mathrm{tr}(\Pi_j\rho)$,
where we have borrowed the double ket (bra) notation from
Refs.~\cite{DarPS00,DarP07}. Furthermore, an out product such as
$\douter{\Pi_j}{\Pi_j}$, which is referred to as an superoperator
henceforth, acts on this space just as an operator acts on the
ordinary Hilbert space (the arithmetics of superoperators can be
found in Refs.~\cite{RunMND00, RunBCH01}). With this background, one
can show that a measurement is IC if and only if the frame
superoperator
\begin{equation}\label{eq:FrameSO1}
\mathcal{F}=\sum_j\frac{\douter{\Pi_j}{\Pi_j}}{\mathrm{tr}(\Pi_j)}
\end{equation}
is invertible \cite{Sco06,DufS52, Cas00, DarP07}. The frame
superoperator $\mathcal{F}$ can be written in the following form
\cite{Sco06}
\begin{eqnarray}
\mathcal{F}=\frac{\mathcal{I}}{d}+\mathcal{F}_0,
\end{eqnarray}
where $\mathcal{I}=\douter{1}{1}$ and
\begin{equation}\label{eq:FrameSOtrless}
\mathcal{F}_0=\sum_j\frac{\douter{\Pi_j-\mathrm{tr}(\Pi_j)/d}{\Pi_j-\mathrm{tr}(\Pi_j)/d}}{\mathrm{tr}(\Pi_j)},
\end{equation}
which is supported on the space of traceless Hermitian operators.
Obviously,  $\mathcal{F}$ is invertible if and only if
$\mathcal{F}_0$ is invertible in this space. In the rest of this
article, $\mathcal{F}_0$ will also be referred to as the frame
superoperator if there is no confusion.

When $\mathcal{F}$ is invertible, there exists a set of
reconstruction operators $\Theta_j$ satisfying
$\sum_j\douter{\Theta_j}{\Pi_j}=\mathbf{I}$, where $\mathbf{I}$ is
the identity superoperator. Given a set of reconstruction operators,
any state can be reconstructed from the set of probabilities $p_j$:
$\rho=\sum_j p_j\Theta_j$. In a realistic scenario, given  $N$
copies of the unknown true state, what we really get in an
experiment are frequencies $f_j$ rather than probabilities $p_j$.
The estimator based on these frequencies $\hat{\rho}=\sum_j
f_j\Theta_j$ is thus different from the true state. Nevertheless,
the deviation $\Delta\rho=\hat{\rho}-\rho$ vanishes in the large-$N$
limit if the measurement is IC. In general, these frequencies obey a
multinomial distribution with the MSE matrix (also called the
covariance matrix)
$\Sigma_{jk}=\bigl(p_j\delta_{jk}-p_jp_k\bigr)/N$. The MSE matrix of
the estimator $\hat{\rho}$ can  be derived according to the
principle of error propagation,
\begin{eqnarray}\label{eq:CovarianceMatrix}
\mathcal{C}(\rho)&=&\sum_{j,k}\dket{\Theta_j}\Sigma_{jk}\dbra{\Theta_k}\nonumber\\
&=&\frac{1}{N}\biggl(\sum_j\dket{\Theta_j}\dinner{\Pi_j}{\rho}\dbra{\Theta_j}-\douter{\rho}{\rho}
 \biggr).
\end{eqnarray}
The MSE is exactly the trace of the MSE matrix,
\begin{eqnarray}\label{eq:MSEg}
\mathcal{E}_{\mathrm{M}}(\rho)&\equiv&\mathrm{E}(\norm{\Delta\rho}^2_{\mathrm{HS}})=\mathrm{Tr}(\mathcal{C}(\rho))\nonumber\\
&=&\frac{1}{N}\biggl(\sum_j
p_j\mathrm{tr}\bigl(\Theta_j^2\bigr)-\mathrm{tr}(\rho^2)\biggr).
\end{eqnarray}
In this article, the symbol ``$\mathrm{Tr}$" is used to denote the
trace for superoperators, and ``$\mathrm{tr}$" that for ordinary
operators. For the convenience of  later discussions, we define
$N\mathcal{E}_{\mathrm{M}}(\rho)$ as the scaled MSE, which is
independent of $N$. The  scaled mean trace distance and  the scaled
mean HS distance can be defined similarly, except that $N$ is
replaced by $\sqrt{N}$.

The set of reconstruction operators   is unique for a minimal IC
measurement, such as a SIC~POM or a product SIC~POM,
 but is not unique for a generic IC measurement. Among all the candidates,
the set of canonical reconstruction operators
\begin{equation}\label{eq:dual}
\dket{\Theta_j}=\frac{\mathcal{F}^{-1}\dket{\Pi_j}}{\mathrm{tr}(\Pi_j)}
\end{equation}
is the best choice for linear state reconstruction in that it
minimizes the MSE averaged over unitarily equivalent true states and
is thus widely used in practice \cite{Sco06}. In the rest of this
article, we will only consider canonical reconstruction operators.
It is then straightforward to verify that $\dket{1}$ is an
eigenvector of $\mathcal{C}(\rho)$ with eigenvalue $0$; in other
words, $\mathcal{C}(\rho)$ is supported on the space of traceless
Hermitian operators as is $\mathcal{F}_0$. The other eigenvalues of
$\mathcal{C}(\rho)$ determine the variances along the principle axes
and thus the shape of the uncertainty ellipsoid.

If $N$ is sufficiently large, the multinomial distribution
approximates  a Gaussian distribution, which  is  completely
determined by the mean and the MSE matrix. In practice, the Gaussian
approximation is already quite good for  moderate values of $N$ if
we are mainly concerned with quantities like the mean HS distance
and the mean trace distance, which are the most common figures of
merit in quantum state tomography. We thus assume the validity of
this approximation in the following discussion. Under Gaussian
approximation, the variance of the squared error
$\norm{\Delta\rho}^2_{\mathrm{HS}}$ is given by the following simple
formula:
\begin{eqnarray}
v(\rho)\equiv\mathrm{Var}(\norm{\Delta\rho}^2_{\mathrm{HS}})=2\mathrm{Tr}\bigl(\mathcal{C}(\rho)^2\bigr).\label{eq:variance}
\end{eqnarray}
In practice, $\sqrt{v(\rho)}$ quantifies the amount of fluctuation
in the squared error  $\norm{\Delta\rho}^2_{\mathrm{HS}}$ over
repeated experiments, that is the typical error in estimating
$\mathcal{E}_{\mathrm{M}}(\rho)$ with just one experiment, assuming
the true state is known.  This error can be reduced by a factor of
$\sqrt{N_{\mathrm{e}}}$ if we repeat the experiment $N_{\mathrm{e}}$
times and take the average of $\norm{\Delta\rho}^2_{\mathrm{HS}}$.
In addition, once $\mathcal{E}_{\mathrm{M}}(\rho)$ is fixed,
$v(\rho)$ also quantifies the dispersion of  the eigenvalues of
$\mathcal{C}(\rho)$, that is the degree of anisotropy  in the
distribution of the estimators.

\subsection{\label{sec:tdesign}Weighted $t$-designs and SIC~POMs}
Consider a weighted set of states $\{|\psi_j\rangle, w_j\}$ with
$0<w_j\leq 1$ and $\sum_j w_j=d$, then the order-$t$ frame potential
$\Phi_t$ for a positive integer $t$ is defined as \cite{Sco06,
RenBSC04}
\begin{eqnarray}\label{eq:t-design}
\Phi_t&=&\sum_{j,k}w_jw_k|\langle
\psi_j|\psi_k\rangle|^{2t}=\mathrm{tr}(S_t^2),\nonumber\\
S_t&=&\sum_{j}w_j(|\psi_j\rangle\langle \psi_j|)^{\otimes t}.
\end{eqnarray}
Note that $S_t$ is supported on the $t$-partite symmetric subspace,
whose dimension is ${d+t-1\choose t}$, $\Phi_t$ is  bounded from
below by $d^2{d+t-1\choose t}^{-1}$ and the bound is saturated if
and only if $S_t=d{d+t-1\choose t}^{-1}P^{\mathrm{sym}}_t$, where
$P^{\mathrm{sym}}_t$ is the projector onto the $t$-partite symmetric
subspace.  The weighted set $\{|\psi_j\rangle,w_j\}$ is  a (complex
projective) \emph{weighted $t$-design} if the lower bound is
saturated; it is a $t$-design if in addition all the weights $w_j$
are equal \cite{Sco06, RenBSC04}. According to the definition, a
weighted $t$-design is also a weighted $t^\prime$-design for
$t^\prime<t$.

It is known that for any positive integers $d$ and $t$, there exists
a (weighted) $t$-design with a finite number of elements
\cite{SeyZ84}. However, the number  is bounded from below by
\cite{Hog82, Sco06}
\begin{eqnarray}\label{eq:tdesignLBound}
{d+\lceil t/2\rceil-1\choose \lceil t/2\rceil}{d+\lfloor
t/2\rfloor-1\choose \lfloor t/2\rfloor},
\end{eqnarray}
which is equal to $d,d^2,d^2(d+1)/2$ for $t=1,2,3$, respectively.
Any resolution of the identity consisting of pure states is a
weighted 1-design. SIC~POMs \cite{Zau11, RenBSC04, ScoG10} and
complete sets of MUB \cite{Iva81, WooF89, DurEBZ10} are prominent
examples of 2-designs. The complete set of MUB for $d=2$ is also a
3-design.

A SIC~POM is composed of $d^2$ subnormalized projectors onto pure
states $\Pi_j=|\psi_j\rangle\langle\psi_j|/d$ with equal pairwise
fidelity \cite{Zau11, RenBSC04, App05, ScoG10}, that is,
\begin{eqnarray}
\label{eq:SICinner}
|\langle\psi_j|\psi_k\rangle|^2=\frac{d\delta_{jk}+1}{d+1},\quad
j,k=1,2\cdots,d^2.
\end{eqnarray}
It is straightforward to verify that a SIC~POM is a 2-design from
this definition. What is not so obvious is that a weighted 2-design
consisting of $d^2$ elements must be a SIC~POM \cite{Sco06}.

A SIC~POM is group covariant if it can be generated from a single
state---the \emph{fiducial state}---under the action of a group
consisting of unitary operators. Most known SIC~POMs are covariant
with respect to the Heisenberg--Weyl  group (also called the
generalized Pauli group) \cite{Zau11, RenBSC04, App05, ScoG10},
which is generated by the phase operator $Z$ and the cyclic shift
operator $X$ defined by their actions on the computational basis,
\begin{eqnarray}\label{eq:HW}
Z|e_r\rangle&=&\omega^r|e_r\rangle, \nonumber\\
X|e_r\rangle&=&\left\{ \begin{array}{cl}
  |e_{r+1}\rangle &\quad\mbox{if}\;  r=0,1,\cdots,d-2, \\
  |e_0\rangle &\quad\mbox{if}\; r=d-1, \\
\end{array}\right.
\end{eqnarray}
where $\omega=e^{2\pi \mathrm{i}/d}$. A fiducial ket $|\psi\rangle$
of the Heisenberg-Weyl group satisfies
\begin{eqnarray}
|\langle\psi|X^{k_1}Z^{k_2}|\psi\rangle|=\frac{1}{\sqrt{d+1}}
\end{eqnarray}
for $k_1,k_2=0,1,\ldots,d-1$ and $(k_1,k_2)\neq (0,0)$. Up to now,
analytical fiducial kets of the Heisenberg-Weyl group are known for
$d\leq16$ and $d=19, 24, 28, 35, 48$ \cite{Zau11, RenBSC04, App05,
ScoG10, Gra11}, numerical fiducial kets with high precision have
been found up to $d=67$ \cite{RenBSC04, RenBSC04b, ScoG10}.

In this  article, all SIC~POMs used in the numerical simulation are
generated by the Heisenberg-Weyl  group from the fiducial kets of
Ref.~\cite{ RenBSC04b}. However, all theoretical analysis is
independent of the specific choice of SIC~POMs.

\subsection{\label{sec:TightIC}Tight IC measurements}
An IC~measurement is \emph{tight}  if the frame superoperator
$\mathcal{F}_0$ is proportional to $\mathbf{I}_0$, that is
$\mathcal{F}_0=a\mathbf{I}_0$ for $a>0$, where $\mathbf{I}_0$ is the
identity superoperator in the space of traceless Hermitian
operators. Scott \cite{Sco06} has shown that the coefficient $a$ is
upper bounded by $1/(d+1)$ for any tight IC measurement, and the
upper bound is saturated if and only if the tight IC~measurement is
rank one. He also showed that rank-one tight IC measurements are
optimal for linear state tomography in the sense that the MSE
$\mathcal{E}_{\mathrm{M}}(\rho)$ averaged over unitarily equivalent
density operators is minimized \cite{Sco06}. Here we shall
recapitulate his main idea in a way that suits our subsequent
discussion.

Since the average of $\rho$ over unitarily equivalent states is the
completely mixed state, according to
Eqs.~(\ref{eq:CovarianceMatrix}) and (\ref{eq:MSEg}), it is enough
to show the optimality of the rank-one tight IC measurements when
the true state is the completely mixed state. In that case, the MSE
matrix and the MSE  can be expressed more concisely in terms of the
frame superoperator $\mathcal{F}_0$,
\begin{eqnarray}\label{eq:CovarianceMatrix2}
\mathcal{C}\Bigl(\frac{1}{d}\Bigr)&=&\frac{1}{N}\Bigl(\frac{\mathcal{F}^{-1}}{d}-\frac{\mathcal{I}}{d^2}\Bigr)=\frac{1}{d
N}\mathcal{F}_0^{-1},\nonumber\\
\mathcal{E}_{\mathrm{M}}\Bigl(\frac{1}{d}\Bigr)&=&\frac{1}{d
N}\mathrm{Tr}\bigl(\mathcal{F}_0^{-1}\bigr).
\end{eqnarray}
The first equation endows  the frame superoperator $\mathcal{F}_0$
with a concrete operational meaning as the inverse of the MSE matrix
(up to a multiplicative factor) evaluated at the point $\rho=1/d$.
From the definitions of the frame superoperators $\mathcal{F}$ and
$\mathcal{F}_0$ (cf. Sec.~\ref{sec:LinearST}), we have
\begin{eqnarray}
\mathrm{Tr}(\mathcal{F}_0)=\mathrm{Tr}(\mathcal{F})-1\leq\sum_j\mathrm{tr}(\Pi_j)-1=
d-1,
\end{eqnarray}
and  the inequality is  saturated if and only if the measurement is
rank one. Recalling that $\mathcal{F}_0$ is supported on the space
of traceless Hermitian operators, whose dimension is $d^2-1$, the
above equation implies that
\begin{eqnarray}
\mathrm{Tr}\bigl(\mathcal{F}_0^{-1}\bigr)&\geq& (d+1)(d^2-1),\nonumber\\
\mathcal{E}_{\mathrm{M}}\Bigl(\frac{1}{d}\Bigr)&\geq&\frac{1}{d
N}(d+1)(d^2-1).
\end{eqnarray}
The inequalities are saturated if and only if
$\mathcal{F}_0=\mathbf{I}_0/(d+1)$.  In other words, rank-one tight
IC measurements are optimal in minimizing the MSE  \cite{Sco06}.

A rank-one tight IC measurement with outcomes $\Pi_j=|\psi_j\rangle
w_j\langle \psi_j|$ features particularly simple canonical
reconstruction operators
\begin{eqnarray}\label{eq:dualTightIC}
 \Theta_j=|\psi_j\rangle(d+1)\langle \psi_j|-1
\end{eqnarray}
and easy state reconstruction.  According to Eq.~(\ref{eq:MSEg}),
the MSE is also given by a simple formula \cite{Sco06}
\begin{eqnarray}\label{eq:MSEtightIC}
\mathcal{E}_{\mathrm{M}}(\rho)=\frac{1}{N}[d^2+d-1-\mathrm{tr}(\rho^2)],
\end{eqnarray}
which is invariant under unitary transformations of the true state.
In addition, the MSE matrix evaluated at $\rho=1/d$ is proportional
to $\mathbf{I}_0$, which means that the uncertainty ellipsoid is
isotropic in the space of traceless Hermitian operators. This
feature will play a crucial  role in our later discussions.

There is a close relation between rank-one tight IC measurements and
weighted 2-designs: A rank-one measurement with outcomes
$\Pi_j=|\psi_j\rangle w_j\langle \psi_j|$ is tight IC if and only if
the weighted set $\{|\psi_j\rangle,w_j\}$ forms a weighted 2-design
 \cite{Sco06}. For example, SIC~POMs and complete sets of MUB are rank-one tight
IC~measurements according to this relation, which can also be
verified directly. Hence, Eqs.~(\ref{eq:dualTightIC}) and
(\ref{eq:MSEtightIC}) are applicable to them. More examples of tight
IC measurements can be found in Ref.~\cite{RoyS07}.

\section{\label{sec:RandomMatrixST}Applications of random matrix theory to quantum state tomography}
In this section, we apply random matrix theory to studying the
tomographic efficiency of tight IC measurements and illustrate the
general result with  SIC~POMs. In particular we derive analytical
formulae for the mean trace distance and the mean HS distance
between the estimator and the true state, thus giving a simple
picture of the resources required to achieve a given accuracy as
quantified  by either of the two distances. Our study also clearly
shows the different scaling behaviors of the two error measures with
the dimension of the Hilbert space. The idea of computing the mean
trace distance using the random matrix theory may also be extended
to derive other figures of merit which only depend on the deviation
between the estimator and the true state.

The rest of this section is organized as follows. In
Sec.~\ref{sec:RandomMatrixIdea}, we present the simple idea of
computing the mean trace distance and the mean HS distance with
random matrix theory. In Sec.~\ref{sec:Isotropic}, we single out
those measurements for which the method is best justified. In
Sec.~\ref{sec:TightICefficiency}, we show that the method works very
well for typical rank-one tight IC~measurements, especially
SIC~POMs. In Sec.~\ref{sec:qubitSIC}, we focus on the qubit SIC~POM.

\subsection{\label{sec:RandomMatrixIdea} A simple idea}

Here is the simple idea of computing the mean trace distance with
random matrix theory: In each experiment, after measurements on $N$
copies of the unknown true state $\rho$, we can construct an
estimator $\hat{\rho}$ for the true state according to the procedure
described in Sec.~\ref{sec:LinearST}.  Once a basis is fixed, the
deviation $\Delta\rho=\hat{\rho}-\rho$ can be represented by a
$d\times d$ matrix, which varies from one experiment to another.
After a large number of repeated experiments, the set of matrices
$\Delta\rho$ can be taken as an ensemble of random matrices obeying
a multidimensional Gaussian distribution, which is completely
determined by  the MSE matrix $\mathcal{C}(\rho)$,
\begin{eqnarray}\label{eq:Gaussian}
p(\Delta\rho)&\propto&\exp\Bigl(-\frac{1}{2}\dbra{\Delta\rho}\mathcal{C}(\rho)^{-1}\dket{\Delta\rho}\Bigr).
\end{eqnarray}
Since $\mathcal{C}(\rho)$ is supported on the space of traceless
Hermitian operators, the distribution of $\Delta\rho$ is restricted
on the hyperplane satisfying $\mathrm{tr}(\Delta\rho)=0$. Suppose
$f(x)$ is the level density function of this ensemble of matrices
with normalization convention $\int \mathrm{d}x f(x)=d$. Then the
mean trace distance between the estimator and the true state is
proportional to the first absolute moment of $f(x)$,
\begin{eqnarray}
\mathcal{E}_{\mathrm{tr}}(\rho)\equiv\frac{1}{2}\mathrm{E}(\mathrm{tr}|\Delta\rho|)=\frac{1}{2}\int
\mathrm{d}x\, |x| f(x).
\end{eqnarray}

If $\mathcal{C}(\rho)$ is (approximately) proportional to the
identity superoperator $\mathbf{I}$, then the ensemble of matrices
$\Delta\rho^\prime=\sqrt{d^2/2\mathcal{E}_{\mathrm{M}}(\rho)}\Delta\rho$
is (approximately) a standard Gaussian unitary ensemble. According
to random matrix theory, for sufficiently large $d$, the level
density $f_{\mathrm{G}}(x)$ of the Gaussian unitary ensemble is
given by the famous Wigner semicircle law \cite{Meh04}:
\begin{eqnarray}
f_{\mathrm{G}}(x)=\left\{\begin{array}{cl}
             \frac{1}{\pi}(2d-x^2)^{1/2}&\quad \mbox{if}\;-\sqrt{2d}\leq x\leq \sqrt{2d},\\
             0 & \quad \textrm{otherwise}.
           \end{array}  \right.
\end{eqnarray}
We can derive $f(x)$ from $f_{\mathrm{G}}(x)$ by a scale
transformation and then compute  the mean trace distance between the
estimator and the true state, with the outcome
\begin{eqnarray}\label{eq:MTRg}
\mathcal{E}_{\mathrm{tr}}(\rho)&\approx&\frac{4}{3\pi}\sqrt{d
\mathcal{E}_{\mathrm{M}}(\rho)}.
\end{eqnarray}
Furthermore, one can verify that the equation is still quite
accurate if $\mathcal{C}(\rho)$ is approximately proportional to
$\mathbf{I}_0$ instead of $\mathbf{I}$, especially when $d$ is
large. In other words, the feasibility of our approach is not
limited by the fact that  $\mathcal{C}(\rho)$ is supported on the
space of traceless Hermitian operators.

When  $\mathcal{C}(\rho)$ is proportional to $\mathbf{I}_0$,
$\Delta\rho$ follows a ($d^2-1$)-dimensional isotropic Gaussian
distribution and $\norm{\Delta\rho}_{\mathrm{HS}}^2$ obeys $\chi^2$
distribution with $d^2-1$ degree of freedom. The mean HS distance
can thus be computed with the result
\begin{eqnarray}\label{eq:MHSg}
\mathcal{E}_{\mathrm{HS}}(\rho)&\equiv&\mathrm{E}\Bigl(\sqrt{\norm{\Delta\rho}^2_{\mathrm{HS}}}\,\Bigr)=\sqrt{\frac{\mathcal{E}_{\mathrm{M}}(\rho)}{d^2-1}}
\frac{\sqrt{2}\Gamma\bigl(\frac{d^2}{2}\bigr)}{\Gamma\bigl(\frac{d^2-1}{2}\bigr)}.
\end{eqnarray}
As a consequence of the law of large numbers, when $d$ is large,
$\mathcal{E}_{\mathrm{HS}}(\rho)$ is approximately equal to the
square root of $\mathcal{E}_{\mathrm{M}}(\rho)$, and with a high
probability the estimator $\hat{\rho}$ is distributed within a thin
spherical shell of radius $\mathcal{E}_{\mathrm{HS}}(\rho)$ that is
centered at the true state.

In general, the accuracy of  Eqs.~(\ref{eq:MTRg}) and (\ref{eq:MHSg})
may depend on the dimension of the Hilbert space and the degree of anisotropy of
the uncertainty ellipsoid  as determined by $\mathcal{C}(\rho)$. However, it turns out that the mean trace distance and the mean HS
distance are not so sensitive to the degree of anisotropy of the
uncertainty ellipsoid. As we shall see shortly, the two equations  are surprisingly accurate for a large family of
measurements, especially tight IC measurements, even if $d$ is very
small (see Fig.~\ref{fig:SICdis}).

Although we have started our analysis with  linear state tomography,
the idea of computing the mean trace distance with random matrix
theory has a wider applicability. We may apply the approach to study
the tomographic efficiencies of other reconstruction schemes, such
as the maximum-likelihood method. In addition, we may also consider
other figures of merit which only depend on the deviation between
the estimator and the true state.

\subsection{\label{sec:Isotropic}Isotropic measurements}
In this section we single out those rank-one IC measurements for
which the uncertainty ellipsoid is the most isotropic,  in which
case Eqs.~(\ref{eq:MTRg}) and (\ref{eq:MHSg}) are best justified.
These measurements turn out to be a special class of tight IC
measurements. In addition to minimizing the MSE, they also minimize
the fluctuation of reconstruction error over repeated experiments.
Moreover, these IC measurements have the nice property that the mean
reconstruction error is almost independent of the true state.

When the true state is the completely mixed state, according to
Sec.~\ref{sec:TightIC}, $\mathcal{C}(1/d)$ is proportional to
$\mathbf{I}_0$ if and only if the measurement is tight IC, and the
coefficient of proportionality is minimized when the measurement is
rank-one. The symmetry requirement on the MSE matrix is thus
 consistent with the efficiency requirement, recall that
rank-one tight IC measurements are optimal for linear state
tomography. Let us focus on the MSE matrix $\mathcal{C}(\rho)$ for a
generic true state, assuming that we have a rank-one tight IC
measurement with outcomes $\Pi_j=|\psi_j\rangle w_j\langle \psi_j|$.
The degree of anisotropy can be quantified by
$\overline{\mathrm{Tr}\bigl(\mathcal{C}(\rho)^2\bigr)}-\overline{[\mathrm{Tr}(\mathcal{C}(\rho))]^2}$,
where the over-line means taking the average over unitarily
equivalent density operators. Since $\mathrm{Tr}(\mathcal{C}(\rho))$
is exactly the MSE, which is the same for all rank-one tight IC
measurements according to Eq.~(\ref{eq:MSEtightIC}), it suffices to
consider $\overline{\mathrm{Tr}\bigl(\mathcal{C}(\rho)^2\bigr)}$.
Note that $\overline{\mathrm{Tr}\bigl(\mathcal{C}(\rho)^2\bigr)}$
also quantifies the fluctuation in
$\norm{\Delta\rho}^2_{\mathrm{HS}}$ over repeated experiments
according to Eq.~(\ref{eq:variance}). We find
\begin{eqnarray}\label{eq:MeanVariance}
&&N^2\overline{\mathrm{Tr}\bigl(\mathcal{C}(\rho)^2\bigr)}=d^2+2d-\frac{2}{d}+\bigl[\mathrm{tr}(\rho^2)\bigr]^2\nonumber\\
&&\quad{}+\frac{(d+1)^3\Phi_3-2(2d^2+3d-1)}{(d-1)}\Bigl[\mathrm{tr}(\rho^2)-\frac{1}{d}\Bigl]\nonumber\\
&&\quad{}-2\Bigl[\frac{2(d+1)}{d+2}\mathrm{tr}(\rho^3)+\frac{d-1}{d+2}\mathrm{tr}(\rho^2)-\frac{1}{d+2}\Bigr]\\
&&\geq d^2+2d-\frac{2}{d}+2\frac{d^2-2}{d+2}\Bigl[\mathrm{tr}(\rho^2)-\frac{1}{d}\Bigl]+\bigl[\mathrm{tr}(\rho^2)\bigr]^2\nonumber\\
&&\quad{}-2\Bigl[\frac{2(d+1)}{d+2}\mathrm{tr}(\rho^3)+\frac{d-1}{d+2}\mathrm{tr}(\rho^2)-\frac{1}{d+2}\Bigr],\quad\label{eq:MeanVarianceLBound}
\end{eqnarray}
where $\Phi_3$ is the order-3 frame potential defined in
Eq.~(\ref{eq:t-design}), and we have used the inequality
$\Phi_3\geq6d/(d+1)(d+2)$ in deriving
Eq.~(\ref{eq:MeanVarianceLBound}). The lower bound is saturated if
and only if $\Phi_3=6d/(d+1)(d+2)$, that is, when the weighted set
$\{|\psi_j\rangle, w_j\}$ forms a weighted 3-design.

An IC measurement derived from a weighted 3-design will be called an
\emph{isotropic measurement} for reasons that will become clear
shortly.  By virtue of the properties of weighted $3$-designs, one
can show that the MSE matrix $\mathcal{C}(\rho)$ is the same for any
IC measurement derived from a weighted 3-design, including the
covariant measurement composed of all pure states weighted by the
Haar measure. In other words, $\mathcal{C}(\rho)$ is invariant under
any unitary transformation of the measurement outcomes:
$\Pi_j\rightarrow U\Pi_jU^\dag$. As an immediate consequence, the
reconstruction error is the same for unitarily equivalent true
states as long as the figure of merit is unitarily invariant, such
as the mean trace distance, the mean HS distance, or the mean
fidelity.

Under linear state tomography,  in addition to  achieving the
minimal MSE $\mathcal{E}_{\mathrm{M}}(\rho)$, an isotropic
measurement also minimizes the fluctuation of the statistical error
over repeated experiments, or equivalently the degree of anisotropy
in the distribution of $\Delta\rho$. One can show that with an
isotropic measurement, $N\mathcal{C}(\rho)$ for a pure true state
has only four (three if $d=2$) distinct eigenvalues, $(d+1)/(d+2),
2(d+1)/(d+2), 0, 2d/(d+2)$ with multiplicities $d(d-2), 2(d-1), 1,
1$, respectively. The degree of anisotropy is even lower if the true
state has a lower purity since the leading term in the expression of
$\mathcal{C}(\rho)$ [cf. Eq.~(\ref{eq:CovarianceMatrix})] is linear
in $\rho$.

In conclusion, Eqs.~(\ref{eq:MTRg}) and ({\ref{eq:MHSg}) are  a good approximation
for computing the mean trace distance and the mean HS distance under isotropic measurements.  After inserting
Eq.~(\ref{eq:MSEtightIC}) into the two equations,  we get
\begin{eqnarray}
\mathcal{E}_{\mathrm{tr}}(\rho)&\approx&\frac{4}{3\pi}\sqrt{\frac{d[d^2+d-1-\mathrm{tr}(\rho^2)]}{N}}
\sim\frac{4}{3\pi}\frac{d^{3/2}}{\sqrt{N}},\label{eq:MTREiso}\\
\mathcal{E}_{\mathrm{HS}}(\rho)&\approx&\sqrt{\frac{d^2+d-1-\mathrm{tr}(\rho^2)}{N(d^2-1)}}
\frac{\sqrt{2}\Gamma\bigl(\frac{d^2}{2}\bigr)}{\Gamma\bigl(\frac{d^2-1}{2}\bigr)}\sim
\frac{d}{\sqrt{N}}.\quad\label{eq:MHSiso}
\end{eqnarray}
The two equations  clearly show the difference in the scaling
behaviors of the two error measures with the dimension of the
Hilbert space.

An isotropic measurement is, in a sense, the most symmetric
measurement allowed by quantum mechanics.  Remarkably, such a
symmetric measurement can be realized with only a finite number of
outcomes and its tomographic efficiency can be characterized by
simple formulae. However, since a weighted 3-design contains at
least $d^2(d+1)/2$ elements [cf. Eq.~({\ref{eq:tdesignLBound})], an
isotropic measurement contains at least $d^2(d+1)/2$ outcomes, which
are much more than the minimum $d^2$ required for an IC measurement.
It is thus of more practical interests to consider generic tight IC
measurements, such as SIC~POMs, which is the focus of the next
section.

\subsection{\label{sec:TightICefficiency}Tight IC POMs and SIC~POMs}
In this section we consider generic rank-one tight IC measurements
\cite{Sco06,RoyS07}, with special emphasis on the minimal tight IC
measurements---SIC POMs \cite{Zau11, RenBSC04, App05, ScoG10}. When
the weighted set $\{|\psi_j\rangle, w_j\}$ forms a weighted 2-design
but not necessarily a weighted 3-design, we can use the inequality
$\Phi_3\leq \Phi_2=2d/(d+1)$ (cf. Sec.~\ref{sec:tdesign}) to derive
an upper bound for $\overline{\mathrm{Tr}(\mathcal{C}(\rho)^2)}$
from Eq.~(\ref{eq:MeanVariance}),
\begin{eqnarray}\label{eq:MeanVarianceUBound}
N^2\overline{\mathrm{Tr}\bigl(\mathcal{C}(\rho)^2\bigr)} \leq
d^2+2d+2d(d+1)\Bigl[\mathrm{tr}(\rho^2)-\frac{1}{d}\Bigl].
\end{eqnarray}
In conjunction with Eqs.~(\ref{eq:variance}) and
(\ref{eq:MSEtightIC}), this equation provides two important pieces
of information. First, the relative deviation
$\sqrt{v(\rho)}/\mathcal{E}_{\mathrm{M}}(\rho)$ is approximately
inversely proportional to $d$; hence,
$\mathcal{E}_{\mathrm{HS}}(\rho)$ is approximately equal to the
square root of $\mathcal{E}_{\mathrm{M}}(\rho)$ and
Eq.~(\ref{eq:MHSiso}) is a good approximation for computing the mean
HS distance, especially when $d$ is large. Second, the degree of
anisotropy in the distribution of $\Delta\rho$ cannot be too high as
long as the measurement is rank-one tight IC. Given that the level
density function $f(x)$ and especially its first absolute moment are
not so sensitive to slight variations in the degree of  anisotropy,
it is reasonable to expect that the mean trace distance  can be
computed approximately by Eq.~(\ref{eq:MTREiso}). This expectation
is supported by extensive numerical simulations.

\begin{figure}

     \includegraphics[width=8cm]{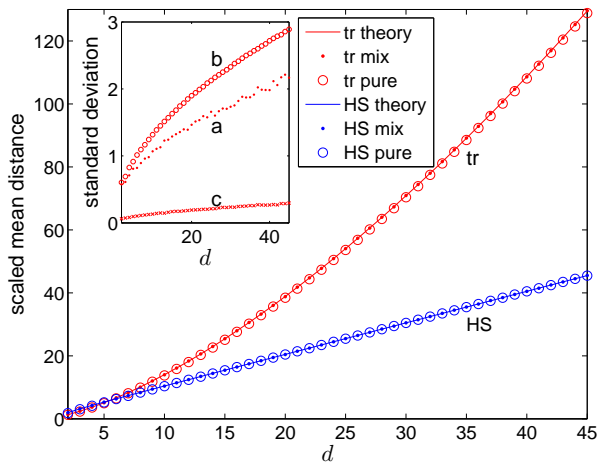}
  \caption{\label{fig:SICdis} (Color online). Theoretical and numerical simulation results
  on the scaled
  mean trace distance $\sqrt{N}\mathcal{E}_{\mathrm{tr}}(\rho)$ and the scaled mean HS distance
  $\sqrt{N}\mathcal{E}_{\mathrm{HS}}(\rho)$ in state tomography with SIC~POMs for $d$ from 2 to 45.
   The SIC POMs are generated by the Heisenberg-Weyl group from the fiducial
  states of Ref. \cite{RenBSC04b}.    The theoretical values are computed according
  to Eqs.~(\ref{eq:MTREiso}) and (\ref{eq:MHSiso}), respectively, with $\rho=1/d$.  $N=1000+20d^2$ is chosen in the numerical simulation. The  values
  for the completely mixed state is the  average over 1000 repeated experiments, and that for pure states is the average over
  1000 randomly-generated pure states each averaged over 100 repeated experiments. The inset shows three kinds of standard deviations of
  the  scaled mean trace distances for the numerical simulation. (a): the standard deviation over repeated experiments
  for the completely mixed state (the jumps in the curve are due to the finite number of experiments);
  (b): the average of the standard deviation over repeated experiments for each pure state;
   (c): the standard deviation over the randomly-generated pure
  states including a partial contribution of the fluctuation over the repeated experiments
for each state due to the finite number of experiments.}
\end{figure}

Figure~\ref{fig:SICdis} shows the results of theoretical calculation
and numerical simulation on state tomography with SIC~POMs. The mean
trace distance and the mean HS distance from numerical simulation
agree perfectly with the theoretical formulae in
Eqs.~(\ref{eq:MTREiso}) and (\ref{eq:MHSiso}), in fact,   much
better than we have expected. The figure also clearly illustrates
the different scaling behaviors of the two error measures with the
dimension of the Hilbert space. From the inset of the figure, we see
that the fluctuation in the mean trace distance over different pure
states is much smaller than the fluctuation over repeated
experiments on the same state. Actually, the former is so small that
it is difficult to separate out the partial contribution of the
latter with a limited number of repeated experiments. In other
words, the reconstruction error is not sensitive to the identity of
the true state.

We emphasize that the results on the tomographic efficiency of
SIC~POMs are representative of typical rank-one tight IC POMs. Since
the order-3 frame potential $\Phi_3=(d^2+3d)/(d+1)^2$ for a SIC~POM
is much larger than the value $6d/(d+1)(d+2)$ required for a
3-design,  a SIC~POM is  a very poor approximation of a 3-design,
for which Eqs.~(\ref{eq:MTREiso}) and (\ref{eq:MHSiso}) are best
justified. Alternatively we can see this  from the value of
$\mathrm{Tr}\bigl(\mathcal{C}(\rho)^2\bigr)$ for a SIC~POM, which
can be computed according to Eqs.~(\ref{eq:CovarianceMatrix}),
\begin{eqnarray}
N^2\mathrm{Tr}\bigl(\mathcal{C}(\rho)^2\bigr)&=&\left(d^2+d+2\right)\bigl[1+\mathrm{tr}(\rho^2)\bigr]-1\nonumber\\
&&{}
+\bigl[\mathrm{tr}(\rho^2)\bigr]^2-2(d^2+d)^2\sum_{j=1}^{d^2}p_j^3.\quad
\end{eqnarray}
When $d\gg1$,  the term $\douter{\rho}{\rho}$ in the expression of
$\mathcal{C}(\rho)$  can be neglected and we have
\begin{eqnarray}
N^2\mathrm{Tr}\bigl(\mathcal{C}(\rho)^2\bigr)&\approx&
(d^2+d)\bigl[1+\mathrm{tr}(\rho^2)\bigr].
\end{eqnarray}
Comparison with Eqs.~(\ref{eq:MeanVarianceLBound}) and
(\ref{eq:MeanVarianceUBound}) shows that the  value
$N^2\overline{\mathrm{Tr}\bigl(\mathcal{C}(\rho)^2\bigr)}$ for a
SIC~POM is roughly in the middle of the lower bound and the upper
bound for tight IC measurements.

In the rest of this section, we briefly examine  tight IC
measurements that are not rank-one and which may arise in practice.
In realistic experiments on quantum state tomography with a SIC~POM,
there always exists noise associated with detector inefficiency,
dark counts, and other imperfections. It is important to understand
how the noise affects tomographic efficiency. We investigate these
effects with a simple white-noise model, in which the  outcomes of
the SIC~POM are modified as follows,
\begin{eqnarray}
\Pi_j(\alpha)=\frac{\alpha
\frac{1}{d}+|\psi_j\rangle\langle\psi_j|}{d\alpha+d},
\end{eqnarray}
where the parameter $\alpha$ ($\alpha\geq0$) characterizes the
strength of the noise. This model is quite natural when there is no
prior knowledge about the noise. Measurements of this form have also
been considered in the context of entanglement detection with
witness operators \cite{ZhuTE10a}.

It is straightforward to verify that the measurement introduced
above is still tight IC. The MSE can be calculated according to the
procedure presented in Sec.~\ref{sec:LinearST}, with the result
\begin{eqnarray}\label{eq:MSEnoise}
\mathcal{E}_{\mathrm{M}}(\rho)&=&\frac{1}{N}\Bigl\{\frac{1}{d}\left[1+(d+1)^2(d-1)(\alpha+1)^2\right]-\mathrm{tr}(\rho^2)
\Bigr\}.\nonumber\\
\end{eqnarray}
Compared with  Eq.~(\ref{eq:MSEtightIC}), the MSE is roughly
$(\alpha+1)^2$ times as large as in the ideal case. The mean trace
distance and the mean HS distance can still be computed according to
Eqs.~(\ref{eq:MTRg}) and (\ref{eq:MHSg}), respectively, with the
result
\begin{eqnarray}
\mathcal{E}_{\mathrm{tr}}(\rho) &\approx&\frac{4}{3\pi
\sqrt{N}}(\alpha+1)d^{3/2},
\nonumber\\
\mathcal{E}_{\mathrm{HS}}(\rho) &\approx&\frac{1}{
\sqrt{N}}(\alpha+1)d, \label{eq:MTREnoise}
\end{eqnarray}
which are roughly $\alpha+1$ times  the values for the ideal case.
Hence, due to the noise, we need roughly $(\alpha+1)^2$ times as
many copies of the true states to reach the same accuracy as in the
ideal case. A similar analysis also applies to tight IC measurements
derived from other 2-designs, such as complete sets of MUB.

\subsection{\label{sec:qubitSIC}Qubit SIC~POM}
In this section we provide further insights on the tomographic
efficiency of the qubit SIC~POM  by deriving an exact formula for
the mean trace distance and discussing the dependence of the
reconstruction error on the Bloch vector of the true state (see
Refs.~\cite{RehEK04, LinLK08, BurLDG08} for earlier accounts).  We
also confirm that the result based on  random matrix theory is
already quite accurate for $d=2$, although it is best justified when
$d$ is large. As a simple application, we  make contact with the
experimental result given by  Ling {\it et al.} \cite{LinLK08}.

For the qubit SIC~POM, the four outcomes $\Pi_k$ for $k=1,2,3,4$ are
in one-to-one correspondence with the four unit vectors
$\boldsymbol{a}_k$ pointing to the four vertices of a regular
tetrahedron inscribed on the Bloch sphere; that is,
$\Pi_k=\frac{1}{4}(1+\boldsymbol{a}_k\cdot\boldsymbol{\tau})$, where
$\boldsymbol{\tau}=(\tau_1, \tau_2, \tau_3)$ are the Pauli matrices
($\sigma_j$ are reserved to denote the standard deviations in this
article). The reconstruction operators are given by
$\Theta_k=\frac{1}{2}(1+3\boldsymbol{a}_k\cdot\boldsymbol{\tau})$
according to Eq.~(\ref{eq:dualTightIC}). Let $\boldsymbol{s}$ denote
the Bloch vector of the true state $\rho$. To reconstruct the true
state $\rho$ is equivalent to  reconstruct the Bloch vector
$\boldsymbol{s}$ \cite{RehEK04},
\begin{eqnarray}
\rho&=&\sum_{k=1}^{4}p_k\Theta_k=\frac{1}{2}\biggl(1+3\sum_{k=1}^4p_k\boldsymbol{a}_k\cdot\boldsymbol{\tau}\biggr),\nonumber\\
\boldsymbol{s}&=&3\sum_{k=1}^4p_k\boldsymbol{a}_k.
\end{eqnarray}
Meanwhile, both the HS  norm $\norm{\Delta\rho}_{\mathrm{HS}}$ and
the trace norm $\norm{\Delta\rho}_{\mathrm{tr}}$ are proportional to
the Euclidean length of
$\Delta\boldsymbol{s}=\hat{\boldsymbol{s}}-\boldsymbol{s}$, where
$\hat{\boldsymbol{s}}$ is an estimator of $\boldsymbol{s}$; that is,
 $\norm{\Delta\rho}^2_{\mathrm{HS}}=(\Delta\boldsymbol{s})^2/2$,
$\norm{\Delta\rho}_{\mathrm{tr}}=|\Delta\boldsymbol{s}|/2$.

The MSE matrix of the estimator $\hat{\rho}$ can be  calculated
according to Eq.(\ref{eq:CovarianceMatrix}), with the result
\begin{eqnarray}
\mathcal{C}(\rho)&=&\frac{3}{4}\biggl(\sum_{j=1}^3\douter{\tau_j}{\tau_j}\biggr)
-\frac{1}{4}\douter{\boldsymbol{s}\cdot\boldsymbol{\tau}}{\boldsymbol{s}\cdot\boldsymbol{\tau}}\nonumber\\
&&{}+\frac{9}{16}\sum_{k=1}^4\dket{\boldsymbol{a}_k\cdot\boldsymbol{\tau}}\boldsymbol{a}_k\cdot\boldsymbol{s}
\dbra{\boldsymbol{a}_k\cdot\boldsymbol{\tau}}.\nonumber
\end{eqnarray}
To get a concrete geometric picture, it is now better to work with
the MSE matrix of the estimator $\hat{s}$ of the Bloch vector,
 \begin{eqnarray}\label{eq:QubitCovarianceM}
\mathcal{C}(\boldsymbol{s})=\frac{1}{N}\Bigl[3\boldsymbol{I}_3-\boldsymbol{s}\boldsymbol{s}
+\frac{9}{4}\sum_k \left(\boldsymbol{a}_k\cdot\boldsymbol{s}\right)
\boldsymbol{a}_k\boldsymbol{a}_k \Bigr],
\end{eqnarray}
where $\boldsymbol{I}_3$ is the $3\times3$ identity dyadic. The mean
squared error of the estimator $\hat{\boldsymbol{s}}$ is given by
\begin{eqnarray}\label{eq:QubitMSE}
\mathrm{E}(|\Delta\boldsymbol{s}|^2)&=&2\mathcal{E}_{\mathrm{M}}(\rho)=\frac{1}{N}(9-s^2).
\end{eqnarray}

Suppose $\sigma_1^2,\sigma_2^2,\sigma_3^2$ are the three eigenvalues
of the MSE matrix $\mathcal{C}(\boldsymbol{s})$, that is the three
variances along the three principal axes of the uncertainty
ellipsoid. Then the mean error is determined by the following
integral,

\begin{eqnarray}
&&\mathrm{E}(|\Delta\boldsymbol{s}|)=\int\mathrm{d}x\mathrm{d}y\mathrm{d}z\frac{\sqrt{x^2+y^2+z^2}}{(2\pi)^{3/2}\sigma_1\sigma_2\sigma_3}
\nonumber\\
&&\hphantom{\mathrm{E}(|\Delta\boldsymbol{s}|)=}\times\exp{\left[-\left(\frac{x^2}{2\sigma_1^2}+\frac{y^2}{2\sigma_2^2}+\frac{z^2}{2\sigma_3^2}\right)\right]}\nonumber\\
&&=\sqrt{\frac{2}{\pi}}\int_0^1\mathrm{d}t\frac{\sigma_1^2\sigma_3^2+\sigma_2^2\sigma_3^2,
+(2\sigma_1^2\sigma_2^2-\sigma_1^2\sigma_3^2-\sigma_2^2\sigma_3^2)t^2}{g^{3/2}},\nonumber
\end{eqnarray}
where
\begin{equation}
g=\sigma_3^2(1-t^2)^2+\frac{\sigma_1^2\sigma_2^2}{\sigma_3^2}t^4
+(\sigma_1^2+\sigma_2^2)t^2(1-t^2).\nonumber
\end{equation}
If at least two of the variances are equal, say $\sigma_2=\sigma_1$,
then the integral can be evaluated explicitly,
\begin{eqnarray}\label{eq:QubitME}
&&\mathrm{E}(|\Delta\boldsymbol{s}|)= \nonumber\\&&\left\{
\begin{array}{cl}
  {\displaystyle\sqrt{\frac{2}{\pi}}\sigma_3} & \mbox{if}\; \sigma_1=0, \\[1ex]
   {\displaystyle\sqrt{\frac{\pi}{2}}\sigma_1} & \mbox{if}\; \sigma_3=0, \\[1ex]
  {\displaystyle2\sqrt{\frac{2}{\pi}}\sigma_1} & \mbox{if}\; \sigma_3=\sigma_1, \\[1ex]
    {\displaystyle\sqrt{\frac{2}{\pi}}\biggl(\frac{\sigma_1^2
    \arctan\sqrt{\frac{\sigma_1^2-\sigma_3^2}{\sigma_3^2}}}{\sqrt{\sigma_1^2-\sigma_3^2}}+\sigma_3  \biggr) }
    & \mbox{if}\; \sigma_1>\sigma_3, \\[2ex]
     {\displaystyle\sqrt{\frac{2}{\pi}}\biggl(\frac{\sigma_1^2
     \mathrm{arctanh}\sqrt{\frac{\sigma_3^2-\sigma_1^2}{\sigma_3^2}}}{\sqrt{\sigma_3^2-\sigma_1^2}}+\sigma_3  \biggr) }
     & \mbox{if}\; \sigma_1<\sigma_3.
\end{array}\right.
\end{eqnarray}

If the uncertainty ellipsoid is
isotropic, that is, $\sigma_1=\sigma_2=\sigma_3$, then we have
\begin{eqnarray}\label{eq:qubitMTRE}
\mathcal{E}_{\mathrm{tr}}(\rho)&=&\sqrt{\frac{2}{3\pi
N}}\sqrt{9-s^2}.
\end{eqnarray}
For the completely mixed state, this equation is exact, by contrast,
the expression
$\mathcal{E}_{\mathrm{tr}}(\rho)\approx4\sqrt{9-s^2}/(3\pi\sqrt{N})$
based on  random matrix theory [see Eq.~(\ref{eq:MTRg})] is about
$8\%$ smaller.  The disparity is much smaller than the relative
deviation of $\norm{\Delta\rho}_{\mathrm{tr}}$ over repeated
experiments, which is about $42\%$. For other true states, the
disparity is even smaller. Hence, the result based on  random matrix
theory is already quite accurate even for $d=2$.
\begin{figure}
   \includegraphics[width=7cm]{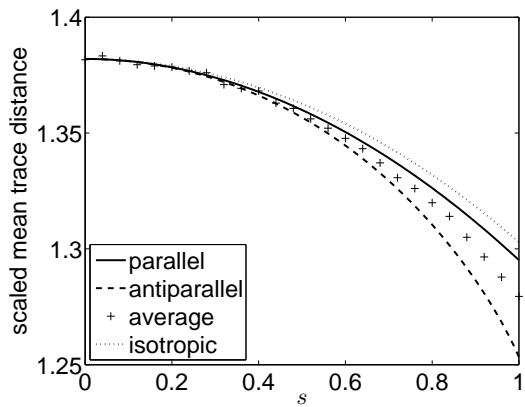}\\
  \caption{The scaled mean trace distances  for states with Bloch vectors that are either parallel
  or antiparallel to the legs of the qubit SIC~POM (theory),
  and the scaled mean trace distance averaged over randomly-generated  states with a fixed purity (numerical
  simulation). $N=1000$ is chosen in  the  numerical simulation, and the scaled trace distance for each given purity is  averaged over
  1000 randomly-generated  states each averaged over 400 repeated
  experiments.
   The dotted curve shows the  scaled mean trace distance determined by Eq.~(\ref{eq:qubitMTRE}),
   which assumes that  the uncertainty ellipsoid is isotropic. }\label{fig:qubitMTE}
\end{figure}

Those states whose Bloch vectors are either parallel or antiparallel
to  the legs of the SIC~POM have attracted more attention both
theoretically \cite{RehEK04} and experimentally \cite{LinLK08},
since they represent two extreme cases. We shall compute the mean
trace distances for those states and discuss the dependence of the
reconstruction error on the Bloch vector of the true state.

If $\boldsymbol{a}_1$ is chosen as the $z$ axis, without loss of
generality, the Bloch vectors of those extreme states can be
parameterized  as $\boldsymbol{s}=z\boldsymbol{a}_1$ with $-1\leq
z\leq 1$. According to Eq.~(\ref{eq:QubitCovarianceM}), the MSE
matrix of $\hat{\boldsymbol{s}}$  now reads
\begin{eqnarray}
\mathcal{C}(s)=\frac{(3-z)\boldsymbol{I}_3+(3z-z^2)\boldsymbol{a}_1\boldsymbol{a}_1}{N},
\end{eqnarray}
whose eigenvalues  are given by
\begin{eqnarray}
\sigma_1^2=\sigma_2^2=\frac{3-z}{N},
\quad\sigma_3^2=\frac{(3-z)(1+z)}{N}.
\end{eqnarray}
Note that the uncertainty ellipsoid is rotationally symmetric. As
$z$ decreases from $1$ to $-1$, the uncertainty ellipsoid evolves
from prolate  to oblate and finally to a singular ellipsoid when
$z=-1$. The mean trace distance of those extreme states can be
calculated according to Eq.~(\ref{eq:QubitME}).
Figure~\ref{fig:qubitMTE} shows the scaled  mean trace distances for
those states together with the numerical average over
randomly-generated states.  The mean trace distance is slightly
smaller for states with Bloch vectors that are antiparallel to the
legs of the SIC~POM than those that are parallel. For a fixed purity
of the true states, the average of the mean trace distance over
randomly-generated states sits roughly  in the middle of the two
extreme cases.  In all three cases, there is a slight decrease in
the mean trace distance as the purity of the true state increases,
which can roughly be attributed to two reasons: the decrease in the
MSEs [cf. Eq.~(\ref{eq:QubitMSE})] and the increase in the degrees
of anisotropy of the uncertainty ellipsoids.

Ling {\it et al.}  \cite{LinLK08} have studied the tomographic
efficiency of the qubit
 SIC~POM experimentally and determined the scaled mean trace distances for the three
 states with $z=0, -1, 1$, respectively, with the result  $1.417, 1.288, 1.323$.
By contrast, our theoretical calculation has yielded the result
$1.382,1.259,1.295$. The experimental and the theoretical values
reflect the same dependence of the reconstruction error on the Bloch
vector of the true state.
  The former are slightly larger than the latter, but the difference is very
  small, in other words, the agreement between experimental data and theoretical calculation is pretty good.
   Note that the relative fluctuation of the reconstruction error over repeated experiments is larger than $40\%$
   and the experimental values are the average of only 40 runs. In addition,
   any imperfection inevitable in real experiments may also  affect the accuracy of the estimator.

\section{Joint SIC~POMs and Product SIC~POMs}\label{sec:sicprod}
In the  bipartite or multipartite settings, it is
technologically much more challenging and sometimes even impossible to perform full joint
measurements such as  SIC~POMs on the whole system. It is thus of paramount practical interests
 to determine  the optimal product measurements and  the efficiency gap between the product measurements and the joint measurements.
We  show that  under linear state tomography, product SIC~POMs are
optimal among all product measurements in the same sense  as joint
SIC~POMs are optimal among all joint measurements. Furthermore, in
the bipartite setting, there is only a marginal efficiency advantage
of the joint SIC~POMs over the product SIC~POMs and it is thus not
worth the trouble to perform the joint measurements. However, in the
multipartite settings, the efficiency advantage of the joint
SIC~POMs over the product SIC~POMs increases exponentially with the
number of parties.

\subsection{Bipartite SIC~POMs and product SIC~POMs}
Consider a product measurement on a bipartite system whose parts
$A,B$ have subsystem dimensions $d_1,d_2$ respectively, and the
total dimension is $d=d_1d_2$.  To show the optimality of the
product SIC~POM, we shall use the same strategy described  in
Sec.~\ref{sec:TightIC}. More generally, we show that  if the product
measurement minimizes the MSE averaged over unitarily equivalent
states, then the measurement on each subsystem is rank-one tight IC,
and vice versa. As an immediate consequence, the product SIC~POM is
optimal and furthermore any minimal optimal product measurement must
be a product SIC~POM, recall that SIC~POMs are the only minimal
tight IC measurements \cite{Sco06}.

Since the average of $\rho$ is the completely mixed state, it
suffices to demonstrate our claim when $\rho=1/d$ according to
Eq.~(\ref{eq:CovarianceMatrix}). Suppose $\Pi_{j_1}$ are the
outcomes of the measurement on the first subsystem and $\Pi_{j_2}$
are those for the second subsystem, then each outcome in the product
measurement has a tensor product form
$\Pi_{j_1j_2}=\Pi_{j_1}\otimes\Pi_{j_2}$. The same is true for the
frame superoperator $\mathcal{F}=\mathcal{F}_1\otimes\mathcal{F}_2$
and the reconstruction operators
$\Theta_{j_1j_2}=\Theta_{j_1}\otimes\Theta_{j_2}$. According to
Eq.~(\ref{eq:CovarianceMatrix2}), we have
\begin{eqnarray}
\mathcal{C}\Bigl(\frac{1}{d}\Bigr)&=&\frac{1}{N}\Bigl(\frac{\mathcal{F}_1^{-1}\otimes\mathcal{F}_2^{-1}
}{d}-\frac{\mathcal{I}}{d^2}\Bigr),\nonumber\\
\mathcal{E}_{\mathrm{M}}\Bigl(\frac{1}{d}\Bigr)&=&\frac{1}{d
N}\bigl[\mathrm{Tr}\bigl(\mathcal{F}_1^{-1}\bigr)\mathrm{Tr}\bigl(\mathcal{F}_2^{-1}\bigr)-1\bigr].
\end{eqnarray}
The MSE is minimized if  and only if both
$\mathrm{Tr}\bigl(\mathcal{F}_1^{-1}\bigr)$ and
$\mathrm{Tr}\bigl(\mathcal{F}_2^{-1}\bigr)$ are  minimized, that is,
when the measurement on each subsystem is rank-one tight IC (cf.
Sec.~\ref{sec:TightIC}).

Next, let us focus on the tomographic efficiency of the optimal
product measurements. If the product measurement is composed of two
rank-one tight IC measurements, as in the case of the product
SIC~POM, then each factor in the reconstruction operator
$\Theta_{j_1j_2}=\Theta_{j_1}\otimes\Theta_{j_2}$ is given by
Eq.~(\ref{eq:dualTightIC}). The MSE can be computed according to
Eq.~(\ref{eq:MSEg}),
\begin{eqnarray}\label{eq:MSEprod}
\mathcal{E}^{\mathrm{prod}}_{\mathrm{M}}(\rho)=\frac{1}{N}\bigl[(d_1^2+d_1-1)(d_2^2+d_2-1)-\mathrm{tr}(\rho^2)\bigr].\quad
\end{eqnarray}
Surprisingly, the MSE is  almost independent of the true state, as
in the case of SIC~POMs. In addition, it is approximately equal to
the product of the MSEs for two subsystems, respectively. The
variance $v^{\mathrm{prod}}(\rho)$ of the squared error may depend
on the specific choice of product measurements according to
Eq.~(\ref{eq:variance}). For the product SIC~POM, it is
approximately given by
\begin{eqnarray}\label{eq:varianceProd}
&&v^{\mathrm{prod}}(\rho)\approx\frac{2}{N^2}\Big\{(d_1^2+d_1-2)(d_2^2+d_2-2)\nonumber\\
&&\quad{}\times\bigl[1+\mathrm{tr}(\rho^2)+\mathrm{tr}(\rho_1^2)+\mathrm{tr}(\rho_2^2)\bigr]+(d_1^2+d_1-2)\nonumber\\
&&\quad{}\times\bigl[1+\mathrm{tr}(\rho_1^2)\bigr]+(d_2^2+d_2-2)\bigl[1+\mathrm{tr}(\rho_2^2)\bigr]
\Big\}.\quad
\end{eqnarray}
Note that the variance does not only depend on the purity of the
global state, but also on the purities of the reduced states, which
means that it generally depends on the entanglement of the global
state. For example, if the true state is pure, the variance is
approximately maximized for product states and minimized for
maximally entangled states.

Compared with the result of the joint SIC~POM given in
Eq.~(\ref{eq:MSEtightIC}), the MSE achieved by the product SIC~POM
is slightly larger,  but the difference is generally very small,
especially when both $d_1$ and $d_2$ are large. By contrast, the
fluctuation over repeated experiments is  stronger by a bigger
margin in the product SIC~POM. Figure \ref{fig:ratio} shows the
ratio of the MSEs when the true state is the completely mixed state,
note that the ratio for other true states is almost the same. The
maximal ratio $1.36$ is obtained when $d_1=d_2=3$. If
$d_1,d_2\geq3$,  the ratio decreases monotonically with $d_1$ and
$d_2$; if $d_2=2$ and $d_1\geq3$, the ratio  decreases monotonically
with $d_1$. For sufficiently large $d_1,d_2$, the ratio is about
$1+1/d_1+1/d_2$. In conclusion, there is only a  marginal
  efficiency advantage of the joint SIC~POM over the product SIC~POM.
  The product SIC~POM is thus more appealing for practical applications since it is much easier to be implemented.

\begin{figure}
        \includegraphics[width=7cm]{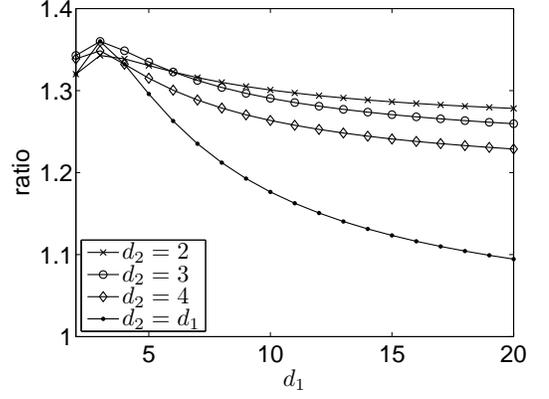}
          \caption{\label{fig:ratio}The ratio of the MSEs in state tomography with the product SIC~POM and
the joint SIC~POM when the true state is the completely mixed state,
note that the ratios for other true states are almost the same. The
ratio is maximized when $d_1=d_2=3$. The ratio of the mean trace
distances is approximately equal to the square root of the ratio of
the MSEs.  }
\end{figure}

Although the product SIC POM is not even a tight IC measurement,
comparison of Eqs.~(\ref{eq:MSEprod}) and (\ref{eq:varianceProd})
shows that the relative deviation of the squared error is quite
small, especially when $d_1,d_2$ are large. Hence,
Eq.~(\ref{eq:MHSg}) is still a good approximation for computing  the
mean HS distance. The mean trace distance can be calculated
approximately according to Eq.~(\ref{eq:MTRg}), with the result
\begin{eqnarray}\label{eq:MTRbiprod}
\mathcal{E}_{\mathrm{tr}}(\rho)
\approx\frac{4\sqrt{d_1d_2}}{3\pi\sqrt{N}}\sqrt{(d_1^2+d_1-1)(d_2^2+d_2-1)-\mathrm{tr}(\rho^2)}.\nonumber\\
\end{eqnarray}
Generally speaking, the larger the values of $d_1$ and $d_2$ are,
the more accurate is this formula. The ratio of the mean trace
distance with the product SIC~POM to that with the joint SIC~POM is
thus approximately equal to the square root of the ratio of the
MSEs.

Table~\ref{tab:sicprod} shows the theoretical and numerical
simulation results on the scaled mean trace distances for the
two-qubit product SIC~POM and joint SIC~POM. There is quite a good
agreement between theoretical calculation and  numerical simulation
although $d_1$ and $d_2$ are so small. The mean trace distances
achieved by the product SIC~POM are roughly $15\%$ larger than that
achieved by the joint SIC~POM. As a consequence, with the product
SIC~POM, we need about $32\%$ more copies of the true states to
reach the same accuracy achieved by the joint SIC~POM. Despite its
slightly lower efficiency, the product SIC~POM is more appealing
than the joint SIC~POM due to the relative ease in its
implementation in real experiments. The same conclusion has also
been reached  in Ref.~\cite{TeoZE10}, where the maximum-likelihood
method was adopted for state reconstruction.

\begin{table}
  \centering
    \caption{\label{tab:sicprod} Theoretical and
numerical simulation results of the scaled mean trace distances
 for the two-qubit  joint SIC~POM
(Joint) and product SIC~POM (Prod). The theoretical values are
computed according to Eqs.~(\ref{eq:MTREiso}) and
(\ref{eq:MTRbiprod}), respectively. $N=1000$ is chosen for the
numerical simulation. For the completely mixed state, the scaled
trace distance is averaged over 1000 repeated experiments. For pure
states, it is averaged over $1000$ randomly-generated states, each
averaged over 1000 repeated experiments. The standard deviations of
the scaled trace distances over the 1000 randomly-generated pure
states (including a partial contribution of the fluctuation over the
repeated experiments for each state due to the finite number of
experiments) are 0.033 and 0.027 for the product SIC~POM and the
joint SIC~POM, respectively, both of which are very small. }
  \begin{tabular}{c|ccc|ccc}
    \hline\hline
   & \multicolumn{3}{|c|}{Completely mixed state}   & \multicolumn{3}{c}{Average over pure states}
    \\
     POM   & Theory & Numerical & Error\% & Theory & Numerical & Error\%\\\hline
    Prod & 4.223  & 4.255 & $-0.8$ & 4.158 & 4.162 & $-0.1$ \\
    Joint &3.676 & 3.716 & $-1.1$ & 3.601 & 3.575 & $+0.7$\\
    Ratio & 1.149 & 1.145 & --- & 1.155 & 1.164 & --- \\
    \hline\hline
  \end{tabular}

\end{table}

\subsection{Multipartite SIC~POMs and product SIC~POMs}
Suppose $k$ parties want to reconstruct a quantum state shared among
them with a product measurement, and $d_j$ for $j=1,2,\cdots,k$ is
the dimension of the Hilbert space of the $j$th party. According to
the same analysis as in the bipartite setting, under linear state
tomography, the product SIC~POM is optimal among all product
measurements. The MSE achieved by the product SIC~POM can also be
calculated in the same manner, with the result
\begin{eqnarray}
\mathcal{E}_{\mathrm{M}}^{\mathrm{prod}}(\rho)=\frac{1}{N}\biggl[\prod_{j=1}^k(d_j^2+d_j-1)-\mathrm{tr}(\rho^2)\biggr],
\end{eqnarray}
which is  almost independent of the true state. When the true state
is the completely mixed state, the variance of the squared error is
given by
\begin{eqnarray}
&&v^{\mathrm{prod}}(\rho)=
\frac{2}{N^2}\biggl[\prod_{j=1}^k\frac{(d_j^3+2d_j^2-2)}{d_j}-\prod_{j=1}^k\frac{1}{d_j^2}
\biggr].
\end{eqnarray}
For a generic true state, the variance  depends on the purity of the
global state as well as  the purities  of various reduced states,
and can be much larger than the value given above.

If the dimension of the  Hilbert space of each party is equal to
$d_1$, then the ratio of the  MSE for the product SIC~POM to that
for the joint SIC~POM grows exponentially with the number of parties
$k$,
\begin{eqnarray}
\frac{\mathcal{E}_{\mathrm{M}}^{\mathrm{prod}}
(\rho)}{\mathcal{E}_{\mathrm{M}}^{\mathrm{joint}}(\rho)}
\approx\left(1+\frac{1}{d}-\frac{1}{d^2}\right)^{-1}\left(1+\frac{1}{d_1}-\frac{1}{d_1^2}\right)^k.
\end{eqnarray}
The ratio of the variances  grows with an even higher rate, whose
specific value may heavily depend on the true state. In other words,
the efficiency advantage of the joint SIC~POM over the product
SIC~POM grows exponentially with the number of parties.

\begin{figure}[t]
              \includegraphics[width=7cm]{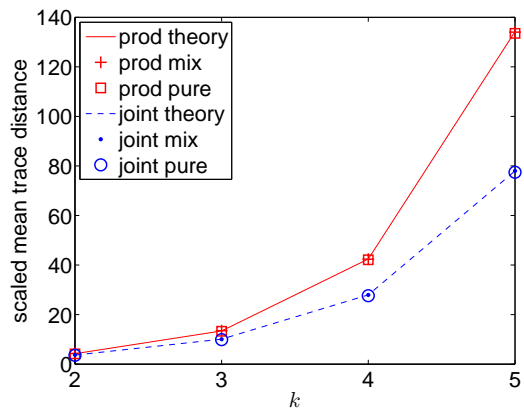}
  \caption{\label{fig:SicMultiProd}(Color online). Theoretical and numerical simulation results of the scaled mean trace distances for the joint
  SIC~POMs  and the product SIC~POMs  on multiqubit systems, where $k$ is the number of qubits. The theoretical values are computed according to
Eqs.~(\ref{eq:MTREiso}) and (\ref{eq:MTRmultiProd}), respectively,
with $\rho=1/d$. $N=1000+20d^2$ is chosen in the numerical
simulation. For the completely mixed state, the scaled  trace
distance is averaged over 1000 repeated experiments. For pure
states, it is averaged over $1000$ randomly-generated states, each
averaged over 100 repeated experiments.
   }
\end{figure}

Although the fluctuation in the reconstruction error over repeated
experiments is stronger in the product SIC~POMs as compared with the
joint SIC~POMs, the relative  fluctuation is still small. Hence,
Eq.~(\ref{eq:MHSg}) is still a good approximation for computing the
mean HS distance. When $k$ is not too large, the mean trace distance
can be calculated approximately according to Eq.~(\ref{eq:MTRg}),
with the result
\begin{eqnarray}\label{eq:MTRmultiProd}
\mathcal{E}_{\mathrm{tr}}(\rho) &\approx&
\frac{4\sqrt{d}}{3\pi\sqrt{N}}\sqrt{\prod_{j=1}^k
\left(d_j^2+d_j-1\right)-\mathrm{tr}(\rho^2)}.
\end{eqnarray}
Since the ratio of the mean trace distance achieved by the product
SIC~POM to that achieved by the joint SIC~POM is approximately equal
to the square root of the ratio of the MSEs, it also increases
exponentially with the number of parties; the same is true for the
mean HS distance. Figure \ref{fig:SicMultiProd} shows theoretical
and numerical simulation results of the scaled mean trace distances
for the product SIC~POMs and the joint SIC~POMs on multiqubit
systems (numerical fiducial kets of SIC~POMs are available at
Ref.~\cite{RenBSC04b}). There is  a pretty good  agreement between
theoretical prediction and numerical simulation for $k$ up to 5.
This plot further confirms  that the efficiency advantage of the
joint SIC~POM over the product SIC~POM increases exponentially with
the number of parties.

\section{Summary}\label{sec:conclusion}
We have introduced random matrix theory \cite{Meh04} for studying
the tomographic efficiency of  tight IC measurements, which include
SIC~POMs as a special example.  In particular, we  derived
analytical formulae for the mean trace distance and the mean HS
distance between the estimator and the true state and showed the
different scaling behaviors of the two error measures with the
dimension of the Hilbert space. The accuracy of these formulae were
confirmed by extensive numerical simulations on state tomography
with SIC~POMs. In the special case of the qubit SIC~POM, we derived
an exact formula for the mean trace distance and discussed in detail
the dependence of the reconstruction error on the Bloch vector of
the unknown true state. As a byproduct, we also discovered  a
special class of tight IC measurements called isotropic
measurements, which feature exceptionally symmetric outcome
statistics and low fluctuation over repeated experiments.

In the bipartite and multipartite settings, we showed that the
product SIC~POMs are optimal among all product measurements in the
same sense as the joint SIC~POMs are optimal among all joint
measurements. We further showed that for bipartite systems, there is
only a marginal efficiency advantage of the joint SIC~POMs over the
product SIC~POMs,  which disappears in the large-dimension limit.
Hence, it is not worth the trouble to perform the joint measurements
at the current stage. However, for multipartite systems, the
efficiency advantage of the joint SIC~POMs over the product SIC~POMs
increases exponentially with the number of parties.

Our study provided a simple picture of the scaling behavior of the
resource requirement in state tomography with the dimension of the
Hilbert space, and of the efficiency gap between product
measurements and joint measurements. The idea of applying  random
matrix theory to studying tomographic efficiencies may also find
wider applications in other state estimation problems.

\section*{Acknowledgements}
We are grateful to Yong Siah Teo for stimulating discussions and
valuable comments on the manuscript. H.Z. would like to thank
Christopher Fuchs and Marcus Appleby for the splendid hospitality at
the Perimeter Institute. This work is supported by NUS Graduate
School (NGS) for Integrative Sciences and Engineering and the Centre
for Quantum Technologies, which is a Research Centre of Excellence
funded by the Ministry of Education and National Research Foundation
of Singapore.

\end{document}